\documentclass[journal,twoside,web]{ieeecolor}
\usepackage{generic}
\usepackage{cite}
\usepackage{amsmath} 
\usepackage{subcaption}
\usepackage{threeparttable}
\usepackage{subcaption}
\newtheorem{definition}{Definition}
\newcommand{\M}{G}

\usepackage{amsmath,amssymb,amsfonts}
\usepackage{graphicx}
\usepackage{algorithm}
\usepackage{hyperref}
\usepackage{textcomp}

\usepackage{algpseudocode}
\newcommand{\G}{\mathcal{G}}

\def\BibTeX{{\rm B\kern-.05em{\sc i\kern-.025em b}\kern-.08em
    T\kern-.1667em\lower.7ex\hbox{E}\kern-.125emX}}
\begin{document}
\title{Privacy-Preserving Personalized Federated Learning (PPP-FL) for Medical Imaging}
\author{Md Morshedul Islam\textsuperscript{*}, 
Hossain Shahriar and Alfredo Cuzzocrea\\
\noindent\textsuperscript{*}Corresponding author: mdmorshedul.islam@concordia.ab.ca
}

\maketitle

\begin{abstract}
Federated learning (FL) is a distributed machine learning paradigm that reduces transmission costs, mitigates cross-border data sharing restrictions, and enhances privacy by keeping raw data on local devices. Clients in FL systems train local models and share only updates, which are aggregated by a central server. While this setup limits direct data exposure, FL remains vulnerable to privacy attacks.
Existing attack and defense strategies have largely been developed for centralized systems, and adapting them to FL introduces new challenges due to its unique architecture and training workflow. In this work, we first propose a model inversion attack on FL systems using Deep Convolutional Generative Adversarial Networks (DCGANs) in the context of large-scale medical imaging. Our method operates under realistic assumptions about the attacker’s knowledge and capabilities. We then examine the effect of client-side differential privacy (DP) in mitigating privacy leakage and incorporate personalized FL to improve model performance under DP settings. Experiments on the Brain MRI ND-5 dataset show that DCGAN-based attacks can reconstruct semantically meaningful synthetic images, revealing significant risks to the privacy of training data. Applying DP reduces attack success but introduces an accuracy trade-off, which is further overcome by personalized FL at each client side. We also evaluate how different aggregation algorithms affect both attack vulnerability and defense effectiveness, underscoring the critical role of aggregation strategies. The proposed attack-and-defense framework is model-agnostic and applicable to a wide range of FL-based systems.
\end{abstract}


%
\IEEEpeerreviewmaketitle

\section{Introduction}
\label{sec:introduction}

The demand for artificial intelligence (AI) has increased significantly over the past decade, driven by advances in machine learning (ML) techniques. Traditionally, ML developers aggregate data on a central server to train a generalized model that can be deployed as a service. However, this centralized approach faces several challenges, including legal restrictions on cross-border data sharing, growing public concerns about personal privacy, and technical difficulties in aggregating large volumes of data from multiple heterogeneous sources. To address these issues, federated learning (FL) has emerged as a promising alternative. FL enables data owners to collaboratively train a global model without sharing raw data, using a central parameter server to coordinate the training. This approach significantly reduces data transmission costs, ensures compliance with data protection regulations, and enhances privacy by design. FL has demonstrated strong potential in privacy-sensitive domains that deal with large-scale data, particularly in healthcare applications such as medical imaging, electronic health record analysis, and disease prediction.

The training process of an FL system begins with a central server constructing a global model and randomly initializing its parameters, which are then distributed to all participating local clients, such as hospitals. Each client performs local training using its private dataset and returns only the local model updates (e.g., gradients or weights), which the server then aggregates using standard algorithms such as Federated Averaging (FedAvg)~\cite{mcmahan2017communication}, resulting in an improved global model. This updated model is then redistributed to the clients for the next training round, and the cycle continues until the model converges or achieves the desired accuracy. While FedAvg remains the most widely used aggregation method due to its simplicity and scalability, it faces challenges in non-IID (non–independent and identically distributed) data and heterogeneous settings, which are common in medical scenarios as well. To address these challenges, several alternative aggregation strategies have been proposed—including Federated Optimization (FedOpt)~\cite{reddi2021adaptive}, Federated Dynamics (FedDyn)~\cite{acar2021feddyn}, and Model-Contrastive Federated Learning (MOON)~\cite{li2021model}—which aim to ensure robust and scalable model training, a key requirement also highlighted in \cite{islam2021scalable}.

The general FL framework is designed to provide a degree of data privacy at scale; however, it remains vulnerable to several privacy threats. Recent studies have shown that various types of privacy attacks can be carried out under different adversarial assumptions, thereby undermining the privacy guarantees of FL systems. Among these, inference attacks~\cite{zhu2019deep, jiang2022comprehensive, zhao2024loki} are particularly concerning, as they may occur during both training and deployment phase. A prominent subclass of inference attacks is model inversion~\cite{islam2021model, hitaj2017deep}, which is widely recognized as one of the most powerful threats to the confidentiality of ML models~\cite{liu2021machine}. Model inversion attacks~\cite{hitaj2017deep, islam2021model} aim to reconstruct class-specific private data used for the training by exploiting the learned relationship between inputs and outputs, often through gradient inversion or surrogate-model optimization. A related category, data reconstruction (or extraction) attacks~\cite{song2020analyzing, zhu2019deep}, focuses on recovering individual training examples without necessarily targeting a specific class. Fredrikson et al.~\cite{fredrikson2015model} first introduced the concept of model inversion by using confidence scores from prediction APIs to reconstruct class-representative inputs for centralized ML. 

There are several motivations for model inversion attacks in FL settings. One practical example in the medical domain occurs when a competitor attempts to extract synthetic, patient-like images from a deployed model to gain access to valuable training data without obtaining regulatory approval or incurring high labeling costs. Such reconstructed data could then be used to train competing models or reverse-engineer proprietary datasets, ultimately undermining the data owner's competitive advantage. While model inversion attacks have been extensively studied in centralized machine learning, extending these attacks to the decentralized and client-centric design of FL introduces unique challenges and remains an active area of research.

This research proposal introduces a Privacy-Preserving Personalized Federated Learning (PPP-FL) framework for medical imaging. We first evaluate the effectiveness of model inversion attacks in FL environments, where an honest-but-curious client colludes with an external adversary to infer sensitive private data. To conduct this attack, we employ a Generative Adversarial Network (GAN)—a deep generative model capable of producing realistic synthetic samples that closely approximate the distribution of the original training data and can successfully deceive the trained global classifier in the FL system. In this setting, the adversary lacks direct access to the original training data or model weights but can interact with the client-deployed trained global model during the inference phase by submitting inputs and observing outputs. The GAN is initially trained on publicly available auxiliary data that are similar in nature to the private training data. Then, by leveraging feedback from the local model the adversary optimizes a latent vector input to the GAN to generate synthetic samples that closely resemble data from a specific target class. For higher-fidelity image-based reconstruction in the model inversion attack, we adopt a Deep Convolutional GAN (DCGAN).

As a defense mechanism, PPP-FL investigates the effect of client-side differential privacy (DP). In this setup, each client applies an DP mechanism by injecting calibrated noise into its model updates before sending them to the server during global model training. This local perturbation obscures sensitive information at the source, thereby reducing the risk of privacy leakage in FL. To implement this approach, we utilize the Opacus privacy engine~\cite{opacus2021}, which provides formal $(\epsilon,\delta)$-DP guarantees at the client level using DP-SGD by adding noise during the training process. Our study also analyzes how different aggregation algorithms affect the vulnerability of FL systems to model inversion attacks and evaluates the effectiveness of client-side DP in mitigating such risks. This dual focus offers valuable information about both algorithmic resilience and privacy protection. We also incorporate personalized FL techniques at the client-level to mitigate the impact of DP on model performance.

The main contributions of this study are as follows:

\begin{itemize}
    \item We propose a DCGAN-based model inversion attack for FL systems, capable of generating synthetic samples that closely resemble the class-specific characteristics of private training data.
    
    \item We investigate the effect of client-side differential privacy (DP) in mitigating model inversion attacks within the FL setting.
    
    \item We analyze how different aggregation algorithms influence both attack success and the effectiveness of DP-based defenses.
    
    \item We incorporate personalized FL techniques at the client-level to alleviate the impact of DP on model performance.
    
    \item We experimentally validate the effectiveness of the proposed attack and defense mechanisms using a large-scale medical MRI dataset.
\end{itemize}

This paper is organized as follows: Section~\ref{sec:background} provides background information, and Section~\ref{sec:related} reviews related work. Section~\ref{sec:attack} presents the proposed attack model, while Section~\ref{sec:dpdefense} describes the integration of client-side DP and personalized FL into the PPP-FL framework and provides a privacy analysis. Section~\ref{sec:experiments} reports the experimental results, highlighting the effectiveness of the attack and the mitigating role of client-side DP and personalized FL. Finally, Section~\ref{sec:conclusion} summarizes key findings and outlines future research directions.

\section{Background} \label{sec:background}
\subsection{Federated Learning}
Google introduced federated learning in 2017 to enable Android users to update models locally without exposing personal data \cite{mcmahan2017communication}. The foundational formulation of the FL system proposed in \cite{mcmahan2017communication} is summarized as follows:
\begin{equation}
    f(\boldsymbol{w})=\sum_{k=1}^K \frac{|D_k|}{|D|} f_k(\boldsymbol{w}),\quad f_k(\boldsymbol{w})=\frac{1}{|D_k|} \sum_{i \in D_k} f_i(\boldsymbol{w}),
    \label{equ1}
\end{equation}
where $f_i(\boldsymbol{w})$ denotes the loss function for the $i$-th training sample $(\boldsymbol{x}_i, y_i)$ parameterized by $\boldsymbol{w}$. Here, $K$ represents the total number of participating clients, and $f_k(\boldsymbol{w})$ is the local objective of client $k$, which contains $|D_k|$ samples stored in its local dataset $D_k$. The total dataset size across all clients is $|D| = \sum_{k=1}^K |D_k|$.  

A common aggregation rule is the Federated Averaging (FedAvg) algorithm~\cite{mcmahan2017communication}, where the global update after each communication round $t$ is computed as follows:
\begin{equation}
    \boldsymbol{w}_{t+1} = \sum_{k=1}^{K} \frac{|D_k|}{|D|} \boldsymbol{w}_t^{(k)},
    \label{equ2}
\end{equation}
where $\boldsymbol{w}_t^{(k)}$ is the locally updated parameter vector of the $k$-th client after completing its local training in round $t$. This weighted averaging ensures that clients with larger datasets contribute proportionally more to the global update.

In a typical FL setting, each local device first receives the current model parameters (weights) from the server and initializes its local model accordingly. The local dataset is then used to update the model by minimizing a local objective function, typically using stochastic gradient descent (SGD). The optimized parameters from all participating devices are sent back to the server, where they are aggregated using an algorithm such as FedAvg. This aggregation step updates the global model, allowing the learning process to continue iteratively. For certain data distributions and system configurations, FL can achieve performance comparable to traditional centralized ML. 

\subsection{Non-Independent and Identically Distributed Data}
In federated learning, non-independent and identically distributed (non-IID) data arises because each client collects data from its own environment, leading to differences in feature distributions, label proportions, or user behaviors. For example, one device may have mostly images of one class while another has completely different classes. This uneven, biased distribution causes local models to learn patterns that do not generalize well globally. As a result, the aggregated model may converge slowly, become unstable, or show degraded performance compared to IID settings.

To model such non-IID distributions in FL simulations, one of the widely adopted strategy is \textit{non-uniform label assignment}~\cite{li2022federated}. Label assignment refers to the way class labels are distributed across clients in FL, which directly influences the degree of data heterogeneity. In our case, by adopting a non-uniform label assignment strategy, we created a controlled non-IID setting. Beyond label imbalance, non-IID data may also arise due to Dirichlet sampling, covariate shift, and concept shift, which are beyond the scope of this paper.

\subsection{Aggregation Algorithm}
FedAvg is one of the most widely used aggregation algorithms in federated learning, but struggle with non-IID data and system heterogeneity. To overcome these challenges, alternatives such as FedOpt~\cite{reddi2021adaptive}, FedDyn~\cite{acar2021feddyn}, and MOON~\cite{li2021model} have been proposed. This paper evaluates these algorithms under model inversion attacks, observing that they do not explicitly incorporate privacy-preserving mechanisms, making them suitable for assessing privacy risks in FL.

FedOpt~\cite{reddi2021adaptive} improves convergence by applying adaptive server-side optimizers such as Adam or Yogi, allowing the global model to better adapt to imbalanced client updates. Despite improving training dynamics, FedOpt may still suffer from fairness concerns and sensitivity to heterogeneous client updates. FedDyn~\cite{acar2021feddyn} introduces a dynamic regularization term to the local objective, which penalizes deviations from the evolving global model. This mechanism mitigates client drift, improving convergence and stability in federated settings with highly skewed data distributions. MOON~\cite{li2021model} further addresses client drift using contrastive learning, encouraging local representations to align with the current global model while diverging from outdated ones. MOON demonstrates strong performance in heterogeneous tasks such as image and text classification and often outperforms FedAvg by improving both consistency and generalization.

\subsection{Personalized Federated Learning}
Traditional FL seeks to learn a single global model $\boldsymbol{w}_g$ that performs well across all clients. However, in the presence of highly heterogeneous data and DP-constrained training settings, a single model often underperforms for certain clients, as it cannot fully capture their local data characteristics. Conversely, training individual local models $\boldsymbol{w}_k$ in isolation may lead to overfitting and poor generalization.

To balance this trade-off, \textit{personalized FL (PFL)}~\cite{huang2021personalized, islam2025privacy} introduces adaptation mechanisms that tailor the global model to each client’s data distribution. The general objective of personalized FL can be conceptually represented as:
\begin{equation}
    \min_{\{\boldsymbol{w}_k,\boldsymbol{w}_g\}}
    \sum_{k=1}^{K} \frac{|D_k|}{|D|} 
    \Big( f_k(\boldsymbol{w}_k) + \lambda \|\boldsymbol{w}_k - \boldsymbol{w}_g\|^2 \Big),
    \label{equ4}
\end{equation}
where $\lambda$ is a regularization parameter that controls the closeness between the local and global models. Smaller $\lambda$ encourages greater personalization, while larger values enforce stronger alignment with the global model.

There are numerous personalization approaches proposed in the literature, such as client clustering, local fine-tuning, model interpolation, and meta-learning-based adaptation. In this work, we adopt local fine-tuning \cite{tan2022towards}, which refers to the process in which, after the global model is trained, each client further trains (fine-tunes) this global model on its own local data to better adapt it to its unique distribution. This technique improves per-client accuracy under highly non-IID settings because the final model is customized to each user's local patterns. Local fine-tuning is one of the simplest and most widely used personalization methods in FL and is particularly helpful for improving performance under DP constraints.

\subsection{Differential Privacy in Federated Learning}
Differential Privacy (DP) is a rigorous framework for protecting individual data contributions while enabling useful aggregate analysis~\cite{dwork2006calibrating,dwork2014algorithmic}. A randomized mechanism $\mathcal{M}$ satisfies $(\epsilon,\delta)$-DP if, for all neighboring datasets $D$ and $D'$ differing by one record and for all measurable output subsets $S$,
\begin{equation}
    \Pr[\mathcal{M}(D) \in S] \leq e^{\epsilon} \Pr[\mathcal{M}(D') \in S] + \delta,
\end{equation}
where $\epsilon$ controls the privacy--utility trade-off and $\delta$ represents a small probability of privacy failure.

DP can be enforced under different trust assumptions. In the \textit{central differential privacy (CDP)} setting, a trusted server aggregates raw data and adds calibrated noise before releasing results. In contrast, \textit{client-side differential privacy (DP)} requires each client to perturb its data locally prior to sharing, thereby providing privacy guarantees even against an untrusted server~\cite{erlingsson2014local}. While true client-side DP typically operates on raw data or locally reported statistics, federated learning more commonly adopts a client-side DP approach in which privacy is enforced during local training.

In federated learning, privacy is typically achieved by applying DP-SGD at the client level, where gradients are clipped and perturbed using Gaussian noise before being sent to the server. This approach limits information leakage from individual training samples while preserving model utility and is more suitable for high-dimensional data. To accurately track cumulative privacy loss across multiple communication rounds, \textit{R\'enyi Differential Privacy (RDP)}~\cite{mironov2017renyi} is commonly used. RDP provides tight composition bounds and is widely supported by DP-SGD implementations such as TensorFlow Privacy and Opacus, enabling a practical balance between privacy protection and learning performance in FL systems.

\begin{table*}[!t]
\footnotesize
\centering
\caption{Summary of representative model inversion attacks.}
\label{tab:model_inversion_related_work}
\renewcommand{\arraystretch}{1.25}
\begin{tabular}{p{2.2cm} p{0.7cm} p{5.2cm} p{3.7cm} p{3.7cm}}
\hline
\textbf{Work} & \textbf{Year} & \textbf{Methodology} & \textbf{Advantages} & \textbf{Limitations} \\
\hline

Fredrikson et al.~\cite{fredrikson2015model} &
2015 &
Exploiting output confidence scores to reconstruct private attributes. &
Demonstrated first attribute inference from model outputs. &
Requires access to model confidence. \\

Hitaj et al.~\cite{hitaj2017deep} &
2017 &
GAN-based collaborative learning attack using shared gradients. &
First GAN-based model inversion attack. &
Not directly applicable to FL settings. \\

Zhu et al.~\cite{zhu2019deep} &
2019 &
Minimizing gradient discrepancy between dummy and shared gradients. &
Works under standard FL gradient-sharing. &
Requires access to the raw gradients.\\

Wang et al.~\cite{wang2019beyond} &
2019 &
GAN-based user-level privacy leakage using multi-task discriminator. &
Reconstructs user-specific samples beyond class-level leakage. &
Requires adversarial server with client-identity knowledge. \\

Zhao et al.~\cite{zhao2020idlg} &
2020 &
Infers true labels from gradients before image reconstruction. &
Removes label ambiguity and converges faster. &
Limited to single-image recovery. \\

Geiping et al.~\cite{geiping2020inverting} &
2020 &
Gradient inversion using cosine similarity and total-variation regularization. &
Produces high-fidelity images on large datasets. &
Depends on the precise gradients. \\

Yue et al.~\cite{yue2021black} &
2021 &
Use autoregressive generation and knowledge distillation. &
Works without direct gradient access. &
Relies on proxy data and domain similarity. \\

Yin et al.~\cite{yin2021see} &
2021 &
Group-consistency regularization for batch-level reconstruction. &
Supports multi-sample image recovery. &
Assumes distinct labels per batch. \\


Zhao et al.~\cite{zhao2022deep} &
2022 &
Use model parameters instead of gradients for inversion. &
Avoids gradient-sharing dependency. &
Requires model parameter snapshots and repeated querying. \\

Xu et al.~\cite{xu2022agic} &
2022 &
Approximate gradient inversion with GAN-like generator–discriminator pair. &
Improves visual fidelity of reconstructed images. &
Requires iterative adversarial training. \\

Yang et al.~\cite{yang2022using} &
2022 &
Reconstructs private data from compressed gradients. &
Shows gradient compression does not ensure privacy. &
Dependent on access to compressed gradient.\\

Liang et al.~\cite{liang2023egia} &
2023 &
External gradient inversion attack using conditional GAN. &
Reconstructs data without altering protocol. &
Needs auxiliary network training. \\

Geng et al.~\cite{geng2023improved} &
2023 &
Gradient inversion with enhanced initialization and regularization. &
Reduces reconstruction loss, improves stability. &
Sensitive to gradient noise and batch sizes. \\

\textbf{Our Approach} &
\textbf{---} &
GAN-based latent-space optimization for FL. &
Operates entirely in loss-based black-box setting. &
Less effective under client-level DP and personalized FL. \\
\hline
\end{tabular}
\end{table*}

\section{Related Work} \label{sec:related}
\subsection{Model Inversion Attack}
In federated learning, model inversion attacks generally proceed in three stages: 
(i) the attacker gains access to a trained global model—either client or server deployed model; 
(ii) the attacker initializes random dummy inputs and iteratively refines them using gradient descent, aiming to minimize the discrepancy between observable model outputs (e.g., gradients, losses, or logits) and those induced by the dummy inputs; and (iii) upon convergence, the attacker obtains synthetic inputs that closely resemble the original private training samples.

A notable early model inversion attack is the work of Fredrikson et al.~\cite{fredrikson2015model}, who demonstrated that confidence scores from model outputs can leak sensitive information, introducing the first formal model inversion attack capable of reconstructing identifiable private attributes. In the domain of recommendation systems, Yue et al.~\cite{yue2021black} introduced a black-box model inversion technique that combines autoregressive generation and knowledge distillation, enabling effective user-data reconstruction even with limited proxy data. 

At the gradient-sharing level, Zhu et al.~\cite{zhu2019deep} proposed \textit{Deep Leakage from Gradients (DLG)}, which reconstructs inputs by minimizing the distance between the dummy and shared gradients using the L-BFGS optimizer~\cite{liu1989limited}. Zhao et al.~\cite{zhao2020idlg} extended this to \textit{iDLG}, directly inferring ground-truth labels before image reconstruction, while Geiping et al.~\cite{geiping2020inverting} incorporated cosine similarity and total-variation regularization to scale reconstruction to large datasets such as ImageNet. Yin et al.~\cite{yin2021see} further proposed \textit{GradInversion}, which enables batch-level image reconstruction via group-consistency regularization. Xu et al.~\cite{xu2022agic} introduced \textit{AGIC}, an approximate gradient inversion framework that integrates a GAN-like generator–discriminator pair to enhance visual fidelity under non-IID conditions, while Liang et al.~\cite{liang2023egia} presented \textit{EGIA}, an external gradient inversion attack that trains a conditional GAN to reconstruct private data from observed model updates without protocol modification. Zhao et al.~\cite{zhao2022deep} and Geng et al.~\cite{geng2023improved} further refined gradient inversion attacks and proposed corresponding defense mechanisms to mitigate the reconstruction quality.

\begin{figure*}[!ht]
    \centering
    \includegraphics[height=6.5cm, width=13cm]{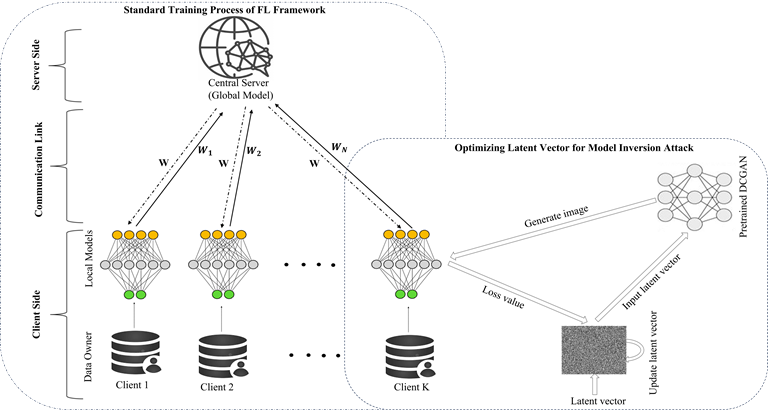}
    \caption{The left panel illustrates how a global model is collaboratively trained across distributed clients without sharing raw data. The right panel shows the DCGAN-based attack, where the generator reconstructs class-specific samples by optimizing latent vectors using feedback obtained from the client-deployed trained global model through loss-based black-box queries.}
    \label{fig:DCGAN-training}
\end{figure*}

Beyond gradient-based techniques, GAN-driven inversion has been explored to amplify privacy leakage. Hitaj et al.~\cite{hitaj2017deep} pioneered a GAN-based collaborative learning attack in which a malicious participant synthesizes class-representative samples from shared gradients. Extending this to federated settings, Wang et al.~\cite{wang2019beyond} demonstrated \textit{user-level privacy leakage} using a multi-task discriminator that jointly predicts class, real/fake labels, and client identity, thereby reconstructing user-specific data. Finally, Yang et al.~\cite{yang2022using} showed that compressing gradients (e.g., via sparsification or quantization) does not effectively prevent reconstruction; their \textit{Highly-Compressed Gradient Leakage Attack (HCGLA)} successfully recovered private data from quantized gradients, disproving the assumption that compression inherently improves privacy.

Our work draws inspiration from Hitaj et al.~\cite{hitaj2017deep}, who proposed a GAN-based model inversion attack in a collaborative learning setting. Their technique leveraged observed gradients to reconstruct samples by inducing class-specific leakage. We extend this approach to FL by assuming an honest-but-curious client that adheres to the protocol while covertly collaborating with an external attacker. Rather than modifying gradients, the client returns model output (e.g., loss values) in response to the attacker’s queries, thereby enabling \textit{loss-based black-box reconstruction} of class-representative data in FL.

\subsection{Defending Against Model Inversion Attack}
To enhance privacy in FL systems—particularly against model inversion attacks—researchers have explored both cryptographic and non-cryptographic techniques. Cryptographic approaches such as Homomorphic Encryption (HE)~\cite{gilad2016cryptonets} and Secure Multi-Party Computation (SMPC)~\cite{bonawitz2017practical} provide strong theoretical privacy guarantees. HE allows computations on encrypted model updates, while SMPC securely aggregates updates without revealing individual contributions. However, their computational and communication costs often limit their practicality for large-scale FL deployments, especially on resource-constrained devices.

To mitigate inference and inversion risks in a more scalable manner, recent works have turned to alternative approaches. Hash-based protocols~\cite{qiu2024hashvfl} and information-theoretic defenses~\cite{tan2024defending} aim to obscure sensitive information without relying on heavyweight encryption. For instance, \textit{HashVFL} applies one-way hashing to local representations, limiting gradient leakage, while information-theoretic frameworks attempt to bound the mutual information between shared model parameters and private data.

Dimensionality reduction~\cite{geiping2020inverting, islam2024privacy, islam2024RPprivacy} and differential privacy (DP)~\cite{abadi2016deep} are two widely adopted non-cryptographic defenses. In FL, dimensionality reduction techniques compress high-dimensional data into compact representations, reducing both communication overhead and the attack surface. Nonetheless, many dimensionality reduction methods lack explicit DP guarantees and may remain vulnerable to sophisticated reconstruction techniques.

Differential privacy, particularly client-side DP implemented via DP-SGD, has shown promise in protecting against gradient-based inversion attacks by injecting noise into updates before they are uploaded to the server~\cite{shokri2015privacy, geyer2017differentially}. Although integrating client-side DP into FL frameworks with various aggregation strategies has shown promise, practical challenges persist due to the trade-off between noise scale and model utility. Recent studies~\cite{huang2021evaluating} have empirically evaluated the effectiveness of client-side DP against inversion attacks and proposed tunable mechanisms to balance the accuracy and privacy. In this work, we incorporate personalized FL techniques together with client-side DP to further balance privacy protection and model utility.

\section{GAN-based Attack Model} \label{sec:attack}
The initial version of our GAN-based model inversion attack for federated learning is accepted for publication in~\cite{islam2025PSBD}. This work extends that study with a more comprehensive analysis of the attack framework, threat model, and privacy risks. We further propose PPP-FL, a defense framework that combines client-side differential privacy and PFL to reduce information leakage while preserving model utility. Figure~\ref{fig:DCGAN-training} overviews the federated learning system and the proposed GAN-based attack, followed by detailed descriptions of the system setting, threat model, and attack and defense methodologies.


\subsection{System Setting}
The FL framework consists of a central server and $K$ clients, forming a cross-silo architecture (e.g., hospitals as clients) in which each client has sufficient computational and data resources to participate in collaborative training. The large-scale dataset is distributed across $n$ classes, but each client holds only a subset of these classes and may lack data for one or more of them. A practical example arises in healthcare, where each client corresponds to a hospital that maintains brain MRI scans for various tumor types. Although an individual hospital may not have samples for certain tumor categories, all hospitals collectively train a global classifier that generalizes across all classes.

Communication between the server and clients is assumed to be secure, for example, through Transport Layer Security (TLS). To further protect against server-side threats and maintain the confidentiality of model parameters during transmission and aggregation, a secure aggregation protocol~\cite{bonawitz2017secagg} can be employed. Once training is complete, the server broadcasts the final global model parameters to all clients, which then deploy the trained global model locally.

\subsection{Threat Model}
In our threat model, the attacker operates in a loss-based black-box setting and collaborates with an honest-but-curious client within the FL system. The client follows the FL protocol but helps the attacker by returning model loss, in response to input queries. The attacker lacks access to the model’s internal architecture, parameters, or training data, but knows the task and loss function used during training. The attacker also possesses a DCGAN that is trained on an auxiliary dataset that shares similar properties to the client’s private data but does not contain samples from the target class. By submitting a limited number of classification queries to the client’s local model, the attacker uses the resulting loss values to iteratively optimize the DCGAN’s latent vector and generate synthetic samples that resemble the target class.

\subsection{Attack Methodology}
The attack methodology consists of four main steps: (i) collecting auxiliary data, (ii) training a DCGAN, (iii) optimizing latent vectors, and (iv) reconstructing the target data. Each step is described in detail below.

\subsubsection{Collecting auxiliary data} 
The attacker is assumed to have general knowledge about the training data used in the federated learning system. For example, the attacker knows the input image dimension, number of channels, and other basic characteristics, which can typically be obtained from an honest-but-curious client. To support the attack, the attacker collects an auxiliary dataset similar in type to the dataset used for training the global model. For instance, if the global model is trained on brain MRI images, the attacker gathers publicly available brain MRI images from online sources. The two datasets-global model training data and auxiliary-are disjoint, meaning they do not share any common samples. Additionally, the auxiliary dataset may be unlabeled if class annotations are not available to the attacker.

\subsubsection{Train a DCGAN}
We used a DCGAN rather than a standard GAN, as DCGANs generate higher-quality synthetic images. The model includes two components: a generator and a discriminator. The attacker trains the generator on an auxiliary dataset to learn a smooth latent representation of the data distribution. Without this generator, the attack would need to optimize directly in the pixel space of the images, which is unstable and typically produces unrealistic results.

{\bf Generator:} The generator in a DCGAN is a deep convolutional neural network that transforms random noise vectors (latent vectors) into synthetic images. It is trained to fool the discriminator by generating images that appear increasingly realistic. During training, the generator is optimized using gradient feedback received from the discriminator, which evaluates how closely the generated images resemble real samples from the auxiliary dataset. In the context of our attack, the generator learns to produce high-quality images that match the distribution of the auxiliary data, even though it has no access to samples from the actual target class.

{\bf Discriminator:} The discriminator in a DCGAN is a convolutional neural network that functions as a binary classifier, distinguishing between real images and fake images generated by the generator. During training, it is optimized using binary cross-entropy loss. Additionally, the discriminator provides gradient feedback to the generator to improve the realism of the generated images. In the context of our attack, the discriminator plays a crucial role in guiding the generator to produce high-quality synthetic images.
    
These two networks are trained in an adversarial setup: the generator learns to produce synthetic images that resemble real data, while the discriminator simultaneously improves its ability to distinguish between real images from the dataset and fake images generated by the generator. 

\begin{algorithm}[!t]
\caption{DCGAN-based Model Inversion Attack.}
\label{alg:dcgan-inversion}
\begin{algorithmic}[1]

\Require 
Pretrained DCGAN generator $\G$; Target class label $c_i$; Black-box access to client-deployed global model $\M$ that returns cross-entropy loss; Latent vector \texttt{$\mathbf{z}$}; step size \texttt{lr}; number of iterations \texttt{T}; number of restarts \texttt{R}; Search mode: \texttt{gradient-free}

\Ensure Set of reconstructed samples \texttt{S}

\State Initialize empty set \texttt{S}

\For{\texttt{each restart r from 1 to R}}
    \State Initialize latent vector $\mathbf{z}$ by sampling from a normal distribution
    \For{\texttt{iteration t = 1 to T}}
        \State Generate image: $\boldsymbol{x} = \G(\mathbf{z})$
        \State Query the model to obtain: \texttt{loss\_0 = $\M(\boldsymbol{x}, c_i)$}
        
        \State Sample a random direction vector \texttt{u}
        \State Compute perturbed latents: 
        \begin{itemize}
            \item \texttt{$\mathbf{z}$\_plus  = $\mathbf{z}$ + epsilon * u}
            \item \texttt{$\mathbf{z}$\_minus = $\mathbf{z}$ - epsilon * u}
        \end{itemize}
        
        \State Query losses for perturbed latents:
        \begin{itemize}
            \item \texttt{loss\_plus  = $\M(\G(\mathbf{z}$\_plus), $c_i$)}
            \item \texttt{loss\_minus = $\M(\G(\mathbf{z}$\_minus), $c_i$)}
        \end{itemize}
        
        \State Estimate search gradient:  
        \State \quad \texttt{grad = (loss\_plus - loss\_minus) / (2 * epsilon) * u}
        
        \State Update latent vector: \texttt{$\mathbf{z}$ = $\mathbf{z}$ - lr * grad}
    \EndFor
    
    \State Add reconstructed image \texttt{$\G(\mathbf{z})$} to the set \texttt{S}
\EndFor

\State \Return \texttt{S}

\end{algorithmic}
\end{algorithm}

\subsubsection{Optimized latent vector} 
Although the generator is trained to produce synthetic images resembling the auxiliary dataset, these samples do not correspond to any specific target class. To reconstruct images of a desired class, the attacker must optimize the latent input of the trained DCGAN.

The process begins with a randomly initialized latent vector $\mathbf{z}$, which is passed through the pretrained generator $\G(\mathbf{z})$ to produce a synthetic image $\boldsymbol{x}$. This image is then submitted to the honest-but-curious client’s local copy of the global model. Using only the model’s output, specifically, the cross-entropy loss between its prediction and the target label, the attacker iteratively updates the latent vector  via gradient-free optimization using finite-difference estimates, steering the generator toward producing images that better align with the target class.

Algorithm~\ref{alg:dcgan-inversion} implements this procedure under a loss-based black-box model-inversion setting, where the attacker interacts with the target model solely through loss queries. For a latent vector $\mathbf{z}$, the generator outputs $\boldsymbol{x} = \G(\mathbf{z})$, and the attacker receives only the loss value $\M(\boldsymbol{x}, c_i)$ from the client's local copy of global model $\M$ for the target class $c_i$, with no access to model gradients or internal parameters. To update $\mathbf{z}$ without gradients, a zeroth-order search is used: at each iteration a random direction $u$ is sampled, two perturbed vectors are generated $\mathbf{z}_{\text{plus}} = \mathbf{z} + \epsilon u$ and $\mathbf{z}_{\text{minus}} = \mathbf{z} - \epsilon u$, and the corresponding losses $\M(\G(\mathbf{z}_{\text{plus}}), c_i)$ and $\M(\G(\mathbf{z}_{\text{minus}}), c_i)$ are queried. These evaluations yield a finite-difference estimate of the search gradient, which is used to update the latent vector.

This optimization loop runs for $T$ iterations and $R$ random restarts, producing a set of reconstructed samples $\G(\mathbf{z})$ at the end of each restart. Because the attacker only observes loss values, the method operates entirely under loss-based black-box constraints. Multiple restarts improve stability and sample quality, increasing the likelihood that the generated images fall within the target class in the model's feature space and are classified by $\M$ as the desired label.

\subsubsection{Target data recovery}
Once the latent vector is optimized, it is used to generate synthetic samples that the target model classifies as valid instances of the target class. By repeating this procedure, the attacker can generate multiple class-consistent images.

The aggregation algorithm used in the FL system plays a pivotal role in the effectiveness of model inversion attacks, as the latent vector optimization process depends on feedback from the model trained under different aggregation strategies. Aggregation methods that promote better generalization across heterogeneous client data tend to yield models with smoother decision boundaries, resulting in more stable loss feedback. This, in turn, enables more accurate finite-difference estimates during optimization, producing synthetic images that more closely resemble the target class. In contrast, weak or noisy feedback--resulting from biased or poorly aggregated models--can hinder optimization and degrade the quality of reconstructed samples while also reducing overall global model performance.

\begin{algorithm}[!ht]
\caption{FedAvg \textit{with Differential Privacy} \em{(FedAvgDP)}.}
\label{alg:fedavgdp}
\begin{algorithmic}[1]
\Require $K$: set of clients; $D$: global dataset; $G$: global model with parameters $w^{(0)}$; 
         $R$: max rounds; $\eta$: learning rate; $\gamma$: client ratio; 
         $E$: local epochs; $\epsilon$: privacy parameter; $\delta$: failure probability parameter
\Ensure Differentially-private, personalized updated global model $G$

\For{$r = 1$ to $R$}
    \State $K_r \gets \text{RandomSample}(K, \gamma |K|)$ \Comment{Select subset of clients}
    \State $L \gets \{\}$ \Comment{Initialize list for local models}
    \For{each client $k \in K_r$}
        \State Initialize $G_k \gets G$
        \Comment{Client receives global model from cloud}
        \State Let $\boldsymbol{w}_k$ be the parameters of $G_k$
        \State Attach PrivacyEngine $(\epsilon, \delta)$ \Comment{Enable DP at client side}
        \For{$t = 1$ to $E$}
            \State Sample batch $(\mathbf{x}, y)$ from local dataset $D_k$
            \State Compute predictions $\hat{y} \gets G_k(\mathbf{x})$
            \State Compute loss $\ell \gets \mathcal{L}(\hat{y}, y)$
            \State Compute per-sample gradients $g_i \gets \nabla \ell_i$
            \State Clip gradients: $g_i \gets g_i / \max(1, \|g_i\|_2 / K_{\max})$
            \State Add Gaussian noise: $\tilde{g} = \frac{1}{|B|}\sum_{i \in B} g_i + \mathcal{N}(0, \sigma^2 K_{\max}^2 I)$
            \State Update local weights: $\boldsymbol{w}_k \gets \boldsymbol{w}_k - \eta \cdot \tilde{g}$
        \EndFor
        \State Apply personalization layer to client $k$’s model \Comment{adapted to local data as in Section~\ref{sec:PersonAlg}}
        \State Evaluate local model $G_k$ and record metrics
        \State $L[k] \gets \boldsymbol{w}_k$
    \EndFor
    \State \textbf{Aggregate:} 
        $\boldsymbol{w}^{(r)} \gets \frac{1}{|K_r|}\sum_{k \in K_r} L[k]$ \Comment{FedAvg aggregation}
    \State Update $G$ with aggregated parameters $\boldsymbol{w}^{(r)}$ and evaluate on global dataset $D$
\EndFor
\State \Return $G$ \Comment{Differentially-private, personalized model}
\end{algorithmic}
\end{algorithm}

\section{Privacy-Preserving Personalized Federated Learning (PPP-FL)} \label{sec:dpdefense}
In PPP-FL, we introduced DP noise on the client side during the training phase of the FL system, thereby enhancing the privacy of client data. To preserve the overall performance of the clients' model, we then incorporated personalized FL.

\subsection{Federated Learning with Client-Side DP}
The training process in PPP-FL is governed by the chosen aggregation algorithm, together with client-side DP. Algorithm~\ref{alg:fedavgdp} outlines the key steps when FedAvg aggregation algorithm is combined with DP in PPP-FL which are: (i) the server randomly selects a subset of $K_r$ clients at each round; (ii) the \textit{PrivacyEngine} adds Gaussian noise to per-sample gradients, controlled by parameters $\epsilon$ and $\delta$; (iii) each selected client performs local training for $E$ epochs on mini-batches; (iv) gradients are clipped, noised, and used to update the local model; (v) clients send their updated model parameters to the server, which aggregates them using FedAvg; and (vi) the updated global model is broadcast to all clients. This workflow is identical when using other aggregation algorithms such as FedOpt, FedDyn, or MOON.

Across all cases, client-side DP is applied during local training according to five key steps: (i) \textit{Per-sample gradient computation:} each client computes gradients for every individual training sample; (ii) \textit{Gradient clipping and accumulation:} gradients whose $L_2$ norm exceeds a predefined threshold are clipped and accumulated; (iii) \textit{Noise addition:} Gaussian noise is injected to obscure individual contributions; (iv) \textit{Gradient scaling:} the noisy gradient is normalized by the expected batch size; and (v) \textit{Optimizer step:} the privacy-preserving gradient is used to update the model parameters.

After completing local training in each round, clients send their differentially private model updates to the central server. This process repeats until the global model converges or reaches the desired accuracy. However, under the non-IID data distributions and DP settings, a single global model may fail to capture client-specific data characteristics and can lead to reduced classification accuracy. Personalized FL addresses this limitation by allowing each client to further adapt the trained global model to its local data.

\subsection{Personalization Algorithm}
\label{sec:PersonAlg}
Personalized federated learning (PFL) of PPP-FL is described in Algorithm~\ref{alg:personalized}. The details of this algorithm are discussed below.


\noindent
\textbf{Model and Objective.}
Each client $k$ holds local non-IID data 
$\{(\boldsymbol{x}_j^{(k)}, y_j^{(k)})\}_{j=1}^{|D_k|}$,
and shares a global model $G: \mathbb{R}^d \to \mathbb{R}^K$ with parameters 
$\boldsymbol{w}_g$, trained collaboratively in the federated and DP settings. During personalization, each client adapts the global parameters $\boldsymbol{w}_g$ to its local data by learning personalized parameters $\boldsymbol{w}_k$, 
minimizing the personalized loss:
\begin{equation}
\mathcal{L}^{PFL}_k(\boldsymbol{w}_k; \boldsymbol{w}_g) 
= \frac{1}{|D_k|} \sum_{j=1}^{|D_k|} 
\ell \big( G(\boldsymbol{x}_j^{(k)}),\, y_j^{(k)} \big)
+ \lambda \|\boldsymbol{w}_k - \boldsymbol{w}_g\|^2,
\label{eq:pfl_obj}
\end{equation}
where $\ell(\cdot,\cdot)$ denotes the cross-entropy loss and $\lambda$ controls 
the trade-off between personalization and global consistency. A smaller $\lambda$ 
emphasizes local adaptation, while a larger $\lambda$ enforces stronger alignment 
with the global model. The cross-entropy loss is:
\vspace{-1mm}
\[
\ell(\hat{y}, y) = -\sum_{i=1}^{K} \mathbf{1}_{\{y=c_i\}} \log \hat{y}_i.
\]
\vspace{-1mm}
The corresponding gradient descent update rule is:
\[
\boldsymbol{w}_k^{(t+1)} 
= \boldsymbol{w}_k^{(t)} 
- \eta \nabla_{\boldsymbol{w}_k^{(t)}}
\Big( 
\mathcal{L}_k^{PPL}(\boldsymbol{w}_k^{(t)}; \boldsymbol{w}_g) 
+ \lambda \|\boldsymbol{w}_k^{(t)} - \boldsymbol{w}_g\|^2 
\Big),
\]
where $\eta$ denotes the local learning rate.

\noindent
\textbf{Accuracy Evaluation.}
Client-level accuracy before and after personalization is computed as:
\vspace{-1mm}
\[
\begin{aligned}
\text{Acc}_{k,\text{before}} &= 
\frac{1}{|D_k^{test}|} \sum_{i=1}^{|D_k^{test}|} 
\mathbf{1}\!\left( \arg\max G(\boldsymbol{x}_i^{(k)}; \boldsymbol{w}_g) 
= y_i^{(k)} \right), \\
\text{Acc}_{k,\text{after}} &= 
\frac{1}{|D_k^{test}|} \sum_{i=1}^{|D_k^{test}|} 
\mathbf{1}\!\left( \arg\max G(\boldsymbol{x}_i^{(k)}; \boldsymbol{w}_k) 
= y_i^{(k)} \right),
\end{aligned}
\]
where $D_k^{test}$ is the number of local test samples.

The global summary metrics are:
\[
\begin{aligned}
\overline{\text{Acc}}_{\text{before}} &= \frac{1}{K} \sum_{k=1}^{K} \text{Acc}_{k,\text{before}}, 
\overline{\text{Acc}}_{\text{after}} &= \frac{1}{K} \sum_{k=1}^{K} \text{Acc}_{k,\text{after}}.
\end{aligned}
\]

\begin{algorithm}[!t]
\caption{Personalized FL with Regularization}
\label{alg:personalized}
\begin{algorithmic}[1]
\Require Global model $G$ with parameters $\boldsymbol{w}_g$; regularization weight $\lambda$; learning rate $\eta$; epochs $E$
\Ensure Personalized models $\{\boldsymbol{w}_k\}_{k=1}^K$, accuracy metrics

\For{each client $k \in K$}
    \State Load $(D_k^{\text{train}}, D_k^{\text{test}})$
    \State Initialize local parameters $\boldsymbol{w}_k \gets \boldsymbol{w}_g$
    \State Evaluate $\text{Acc}_{k,\text{before}} \gets \text{Evaluate}(\boldsymbol{w}_k, D_k^{\text{test}})$

    \For{epoch $=1$ to $E$}
        \For{each minibatch $(\boldsymbol{x}, y) \in D_k^{\text{train}}$}
            \State Compute prediction $\hat{y} = G(\boldsymbol{x}; \boldsymbol{w}_k)$
            \State Compute empirical loss $\mathcal{L}_k = \ell(\hat{y}, y)$

            \State Compute regularization loss:
            \[
            \mathcal{L}_{\text{reg}} = \lambda \|\boldsymbol{w}_k - \boldsymbol{w}_g\|^2
            \]

            \State Total personalized loss: $\mathcal{L}_k^{\text{PFL}} = \mathcal{L}_k + \mathcal{L}_{\text{reg}}$

            \State Update personalized parameters:
            \[
            \boldsymbol{w}_k \gets \boldsymbol{w}_k - \eta \nabla_{\boldsymbol{w}_k}\mathcal{L}_k^{\text{PFL}}
            \]
        \EndFor
    \EndFor

    \State Evaluate $\text{Acc}_{k,\text{after}} \gets \text{Evaluate}(\boldsymbol{w}_k, D_k^{\text{test}})$

    \State Store metrics: $\text{metrics}[k] \gets (\text{Acc}_{k,\text{before}},\text{Acc}_{k,\text{after}})$
\EndFor

\State Compute global averages:
\[
\overline{\text{Acc}}_{\text{before}} = \frac{1}{K} \sum_{k=1}^{K} \text{Acc}_{k,\text{before}}, 
\overline{\text{Acc}}_{\text{after}} = \frac{1}{K} \sum_{k=1}^{K} \text{Acc}_{k,\text{after}}
\]

\State \Return $\{\boldsymbol{w}_k\}$, metrics, and averaged accuracies
\end{algorithmic}
\end{algorithm}

\section{Security and Privacy Analysis}
PPP-FL is designed to preserve the confidentiality of clients’ private training data. This section presents a detailed analysis of the system’s privacy and security properties, with particular emphasis on its resilience to information leakage and adversarial inference attacks.

\subsection{Privacy Analysis}
In the context of model inversion attacks, the objective of the adversary is to reconstruct synthetic samples that closely resemble the class-specific private training data. The effectiveness of such attacks can be evaluated by measuring the similarity between the reconstructed synthetic samples and the corresponding private data, typically through a comparison of their feature representations.

\begin{definition}
Consider an attacker attempting a model inversion attack on a federated learning (FL) system to reconstruct private images from a set \( \boldsymbol{X} \) belonging to a class \( c \), resulting in synthetic reconstructions \( \boldsymbol{X}' \). The FL system is said to satisfy \( (1 - \eta) \)-privacy against model inversion attacks if the expected cosine similarity between the feature vector of any synthetic image \( \boldsymbol{x}' \in \boldsymbol{X}' \) and any original image \( \boldsymbol{x} \in \boldsymbol{X} \) for any target class \( c_i \) is less than or equal to a threshold \( \eta \). In other words, the reconstructed images in \( \boldsymbol{X}' \) must not closely resemble the private images in \( \boldsymbol{X} \) in the feature space, ensuring that the system preserves privacy by limiting the degree of similarity between synthetic and original images.
\end{definition}

There are various ways to measure the closeness between two data instances in the feature space. One of the most commonly used methods is \textit{cosine similarity}, which quantifies the similarity between two feature vectors by computing the cosine of the angle between them. The similarity score of cosine similarity ranges from -1 to 1. However, in many practical applications-such as image embeddings-the values typically lie between 0 and 1, where 1 indicates perfect similarity and 0 implies orthogonality or unrelated.

{\bf Privacy evaluation.} We estimate the cosine similarity between two images using the trained global model $G$ of the FL system. Because this model is trained on diverse client datasets, it is capable of capturing meaningful feature representations. Given that the attacker aims to generate synthetic data that closely resembles real data, employing global model for feature extraction from both real and synthetic images offers a realistic and relevant basis for assessing potential privacy risks.

To extract meaningful features from input image, we use only the convolutional layers of the trained global model. Let $\phi(\cdot)$ denote the feature extractor formed by the convolutional layers of the trained global model. This allows us to obtain feature vector $f_{\boldsymbol{x}}=\phi(\boldsymbol{x})$ for the real image $\boldsymbol{x}$ and 
$f_{\boldsymbol{x}'}=\phi(\boldsymbol{x}')$ for the reconstructed synthetic image $\boldsymbol{x}'$. We then compute the cosine similarity $\mathrm{CosSim}(f_{\boldsymbol{x}}, f_{\boldsymbol{x}'})$ between $f_{\boldsymbol{x}}$ and 
$f_{\boldsymbol{x}'}$ to assess how closely the reconstructed image resembles the original in the model’s feature space. For a list of such similarity values computed across multiple reconstructions from $\boldsymbol{X}$ and $\boldsymbol{X}'$, the system is said to provide $(1 - \eta)$-privacy if the maximum similarity value does not exceed the threshold $\eta$, indicating that no reconstructed image is more than $\eta$ close (similar) to any real image in the feature space.

\subsection{Security Analysis}
{\bf Mitigation via differential privacy.} DP noise added at the client end of PPP-FL limiting gradient leakage under bounded adversarial assumptions, thereby reducing the risk of gradient-based inference attacks. Any resulting accuracy trade-offs can be managed through adaptive noise calibration or hybrid privacy schemes~\cite{geyer2017differentially, bonawitz2017practical, abdel2024privacy}. Rather than implementing full adversarial testing or a detailed threat model, we focus specifically on differential privacy as a lightweight mechanism for mitigating gradient leakage risks in FL. Attacks that fall outside the protection offered by DP-such as stronger active adversaries or cryptographic-level threats-are beyond the scope of this work.

{\bf Robustness through personalized FL.} Personalized FL in PPP-FL fine-tunes the global model to each client’s local data and the regularization parameter. This limits the impact of poisoned updates, accelerates convergence under non-IID data and DP settings.

\section{Experimental Results} \label{sec:experiments}

{\bf Dataset:} We implemented and evaluated PPP-FL using {\em Brain MRI ND-5} dataset ~\cite{tumordata}, a large-scale benchmark hosted on IEEE DataPort and widely adopted for brain tumor classification research. The dataset contains MRI images from four categories: \textit{glioma}, \textit{meningioma}, \textit{pituitary}, and \textit{no\_tumor}, making it well-suited for multi-class classification tasks. In total, it provides 13{,}927 training, and 3{,}961 testing images in RGB format, with resolutions ranging from $256$ to $512$ pixels. While large-scale, the dataset is imbalanced across classes, with training image counts varying between 2{,}339 and 4{,}484 per class.

{\bf Data preprocessing:} 
To efficiently handle the large-scale MRI dataset and facilitate scalable training, all images were resized to $64 \times 64$ pixels and converted from RGB to grayscale. This preprocessing step significantly reduces input dimensionality while retaining the critical structural patterns of the scans, thereby lowering computational overhead and expediting the federated learning training process. 

To ensure robust learning on this large-scale dataset and mitigate class imbalance, we applied balancing techniques to equalize each category to 2,500 MRI images. For underrepresented classes, synthetic samples were generated through data augmentation operations, including random flipping, rotation, and brightness adjustment. For overrepresented classes, random downsampling was applied to align with the target size. These balancing strategies not only reduce training bias but also enhance the scalability and generalization capability of the model when processing large volumes of data.

To verify that preprocessing did not affect performance, we trained centralized ML models on the preprocessed dataset. A customized CNN classifier achieved 97.45\% training accuracy within 100 epochs and 96.61\% test accuracy, confirming the effectiveness of the pipeline.

\begin{table}[!b]
\centering
\caption{The global model in the PPP-FL setup is a CNN with convolutional and fully connected layers.}
\begin{tabular}{p{3.0cm}p{2.2cm}p{1.8cm}}
    \hline
    \textbf{Layer (type)} & \textbf{Output Shape} & \textbf{Param \#} \\ \hline
    Conv2D (Conv1)              & (32, H, W)         & 320 \\
    MaxPooling2D (Pool1)        & (32, H/2, W/2)     & 0 \\
    Conv2D (Conv2)              & (64, H/2, W/2)     & 18,496 \\
    MaxPooling2D (Pool2)        & (64, H/4, W/4)     & 0 \\
    Conv2D (Conv3)              & (64, H/4, W/4)     & 36,928 \\
    MaxPooling2D (Pool3)        & (64, H/8, W/8)     & 0 \\
    Flatten                     & (64 × M × M)       & 0 \\
    Dense (FC1)                 & (128)              & depends on M \\
    ReLU Activation             & (128)              & 0 \\
    Dense (FC2)                 & ($n$) & depends on $n$ \\
    \hline
    \multicolumn{3}{p{7.5cm}}{\footnotesize \textbf{Note:} $H$, $W$ denote input height and width; $M$ is the spatial size after pooling; $n$ is the number of classes.}
\end{tabular}
\label{tab:globalmodel}
\end{table}

\begin{figure*}[!t]
\centering
\subfloat[Without DP\label{fig:withoutLDP}]{%
  \includegraphics[width=8cm, height=6cm]{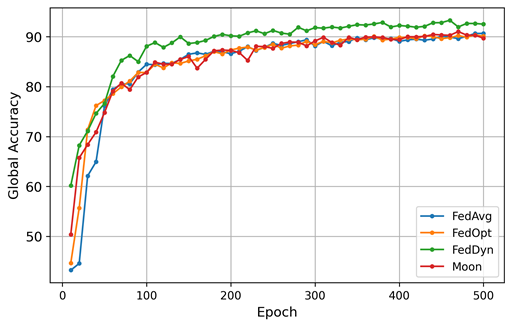}%
}\hfil
\subfloat[With client-side DP\label{fig:withLDP}]{%
  \includegraphics[width=8cm, height=6cm]{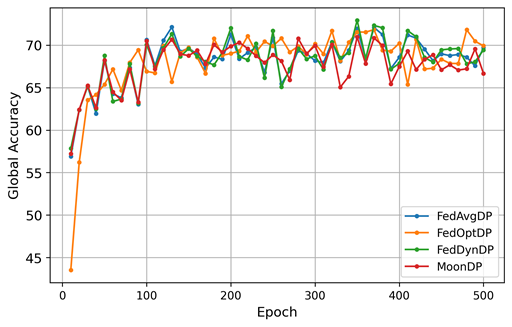}%
}
\caption{Global model accuracy across all four aggregation algorithms of PPP-FL. Non-DP models exceed 90.0\% accuracy after 500 epochs, while client-side DP-enabled models converge more slowly and fluctuate, reaching around 70.0\% accuracy.}
\label{fig:accuracy-loss}
\end{figure*}

{\bf Experiment setup:} In PPP-FL, we adopted a \textit{non-uniform label assignment} strategy to create a controlled non-IID data setting. Label assignment refers to the way class labels are distributed across the clients in FL, which directly influences the degree of data heterogeneity. In our case, the training data from the {\em Brain MRI ND-5} dataset was divided among 10 clients such that each client received samples from a maximum of three of the four classes, meaning each client lacked one or more classes entirely. Within the classes assigned to a client, samples were allocated evenly to maintain balance. This setup preserves the global class distribution while enforcing systematic class differences across clients. Using these partitions, the clients collaboratively trained a global model under different FL aggregation algorithms. Note that the data balancing performed during preprocessing is applied only to the global dataset and does not change the non-IID nature of the client partitions. Each client still receives only a subset of classes during label assignment, so the intended class-skewed (non-IID) distribution is fully preserved.

The test data from the \emph{Brain MRI ND-5} dataset was used as auxiliary data to support the model inversion attack. To simulate an unlabeled setting for DCGAN training, labels were removed from the auxiliary samples. A DCGAN was then trained on this unlabeled data, and latent vectors were subsequently optimized using feedback from client-deployed trained global model under both client-side DP and non-DP configurations. This setup allowed us to assess the privacy vulnerabilities of the FL system in both privacy-preserving and standard training scenarios. During FL training, each client updates its local model, optionally incorporating PFL, and the server aggregates these updates to obtain the global model. A detailed description of each step is provided below.

\subsection{Federated Learning on MRI Dataset}
This section evaluates the performance of PPP-FL systems under both non-DP and DP settings. This analysis is crucial because the success of the model inversion attack depends on the performance of the global model and the degree of its generalization.

\subsubsection{Global model architecture} \label{garchitecture}
For the global model of PPP-FL, each client employs a custom convolutional neural network (CNN) tailored for MRI classification. The model accepts \(64 \times 64\) single-channel images and processes them through three \texttt{Conv2D} layers, each followed by \texttt{ReLU} activation and \texttt{MaxPooling2D}, to extract features while reducing spatial dimensions. Each convolution layer uses $3\times3$ filters with stride $(1,1)$ and padding $(1,1)$ to preserve the input size. The final feature map is flattened into a vector of size \(M=(64\times 8 \times 8)\), followed by two \texttt{Dense} layers with a \texttt{ReLU} activation in between. The final layer outputs logits for \(n=4\) target classes. This architecture remains consistent across all global models; only the aggregation algorithms differ, both in the presence and absence of the DP. Details of the global model architecture shown in Table~\ref{tab:globalmodel}.

\subsubsection{Performance of the global model without DP}
During training, all clients participated in every global training round to maintain consistent model updates. In each round, every client performed 5 local training epochs before sending its updated model parameters to the central server. The server then aggregated the received parameters and redistributed the updated global model to all clients. This process was repeated for 500 global communication rounds, allowing the global model to gradually converge to a satisfactory level of accuracy.

Figure~\ref{fig:withoutLDP} illustrates the accuracy of the global model of PPP-FL under four different aggregation algorithms in the non-DP setting. FedAvg, implemented with SGD and a learning rate of \( lr = 0.03 \), shows the slowest initial convergence-starting with the lowest accuracy. However, it gradually improves, reaching a final accuracy close to 90\%, despite trailing the other methods early in training. FedOpt, also based on SGD but with a higher learning rate of \( lr = 0.1 \), converges faster than FedAvg. Its accuracy increases rapidly in the initial epochs and stabilizes near 90\%, reflecting consistent performance. MOON, trained with \( lr = 0.03 \) and \( \lambda = 1 \times 10^{-3} \), performs similarly to FedAvg and FedOpt. It shows a steady increase in accuracy and reaches nearly 90\% at the end of 500 rounds. FedDyn, using \( lr = 0.03 \) and a regularization parameter \( \lambda = 1 \times 10^{-3} \), achieves the highest accuracy, exceeding 92\%, and demonstrates stable convergence throughout training. Although the final accuracies of all four methods are comparable, their convergence patterns differ notably in the early stages of training before stabilizing over time.

\subsubsection{Performance of the global model with DP}
To train the global model with client-side DP, we used the same model architecture, hyper-parameters, and number of training rounds as used in the non-DP setting. During training, each client uses Opacus’s \texttt{PrivacyEngine.make\_private\_with\_epsilon()} to apply client-side DP to the model, optimizer, and \texttt{DataLoader}. This process clips each individual gradient to a fixed norm (\texttt{max\_grad\_norm}) and adds calibrated noise (\texttt{noise\_multiplier}) to the aggregated gradients. The noisy gradients are then used to update the model, ensuring formal DP guarantees under the specified privacy budget.

Figure~\ref{fig:withLDP} illustrates the accuracy of the global model of proposed PPP-FL for four aggregation algorithms when DP is applied with \textbf{$\epsilon = 5.0$} and \mbox{\textbf{$\delta = 10^{-5}$}} for moderate privacy protection. FedAvg starts at 56.89\% accuracy and gradually increases to nearly 70\%, reflecting instability due to noise injection. FedOpt reaches around 72\% accuracy mid-training but shows periodic drops, highlighting the trade-off between faster convergence and sensitivity to DP noise. FedDyn also approaches 72\%, though its final accuracy slightly declines toward the end of training. MOON tracks closely behind FedDyn, achieving approximately 71\% with greater stability. Overall, while the final accuracies across all four methods are relatively close, their training dynamics under DP vary slightly. As expected, all models exhibit reduced performance compared to the non-DP setting for this complex brain MRI dataset. The tuning of $\epsilon$ directly influences model accuracy, where stronger privacy (smaller $\epsilon$) results in lower accuracy and weaker privacy (larger $\epsilon$) yields higher accuracy, though performance always remains below that of the non-DP setting.

\begin{table}[!t]
\centering
\caption{DCGAN generator architecture detailing the sequence of layers, their output shapes, and applied functions.}
\begin{tabular}{p{2.7cm}p{1.6cm}p{2.7cm}}
    \hline
    \textbf{Layer (type)} & \textbf{Output Shape} & \textbf{Activation / Norm} \\
    \hline
    Input (latent vector $z$)              & (1024, 4, 4)     & ReLU + InstanceNorm \\
    ConvTranspose2D (4×4, stride=2, pad=1)        & (512, 8, 8)      & ReLU + InstanceNorm \\
    ConvTranspose2D (4×4, stride=2, pad=1)        & (256, 16, 16)    & ReLU + InstanceNorm \\
    ConvTranspose2D (4×4, stride=2, pad=1)        & (128, 32, 32)    & ReLU + InstanceNorm \\
    ConvTranspose2D (4×4, stride=2, pad=1)        & (1, 64, 64)      & Tanh \\
    \hline
\end{tabular}
\label{tab:generator}
\end{table}

\begin{table}[!t]
\centering
\caption{DCGAN discriminator architecture detailing the sequence of layers, output shapes, and applied functions.}
\begin{tabular}{p{2.6cm}p{1.5cm}p{3.0cm}}
    \hline
    \textbf{Layer (type)} & \textbf{Output Shape} & \textbf{Activation / Norm} \\
    \hline
    Input Image\\$(1, 64, 64)$              & (64, 32, 32)     & LeakyReLU (0.2) \\
    Conv2D\\(4×4, stride=2)                 & (128, 16, 16)    & LeakyReLU + BatchNorm \\
    Conv2D\\(4×4, stride=2)                 & (256, 8, 8)      & LeakyReLU + BatchNorm \\
    Conv2D\\(4×4, stride=2)                 & (512, 4, 4)      & LeakyReLU + BatchNorm \\
    Conv2D\\(4×4, stride=2, pad=0)          & (1, 1, 1)        & Sigmoid \\
    \hline
\end{tabular}
\label{tab:discriminator}
\end{table}

\subsection{DCGAN-based Attack} 
This section provides a detailed overview of the DCGAN framework, including the architectures of the generator and discriminator, generator's training process with auxiliary data, and the procedure for optimizing latent vectors to recover class-specific images under both non-DP and DP settings.

\begin{figure}
\centering
\subfloat[Original images\label{fig:MRI}]{%
  \includegraphics[width=8cm, height=5.5cm]{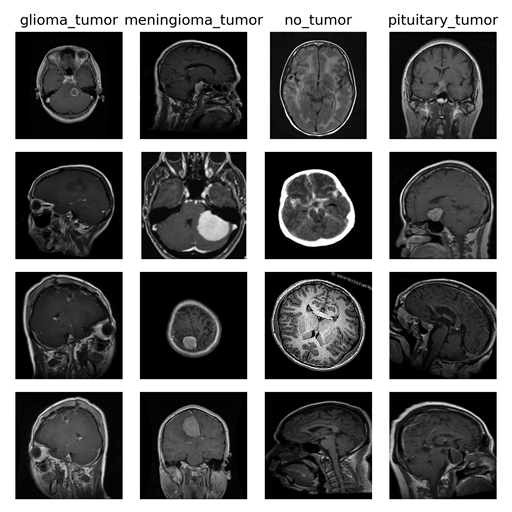}
}\hfil
\subfloat[Synthetic images\label{fig:SynMRI}]{%
  \includegraphics[width=8cm, height=5.5cm]{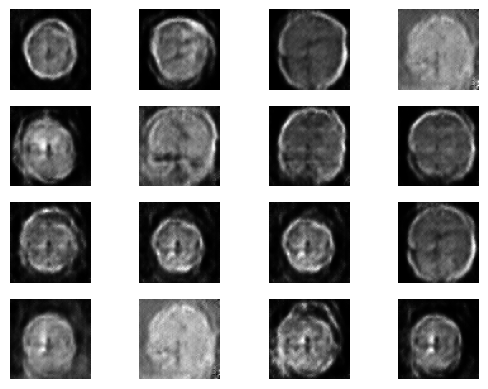}
}
\caption{Comparison of grayscale brain MRI images and synthetic images generated by the trained DCGAN generator using auxiliary data.} \label{fig:origfakeimage}
\end{figure}

\begin{figure*}[!ht]
\centering
\subfloat[non-DP\label{fig:withoutLDPLV}]{%
  \includegraphics[width=8cm, height=5.5cm]{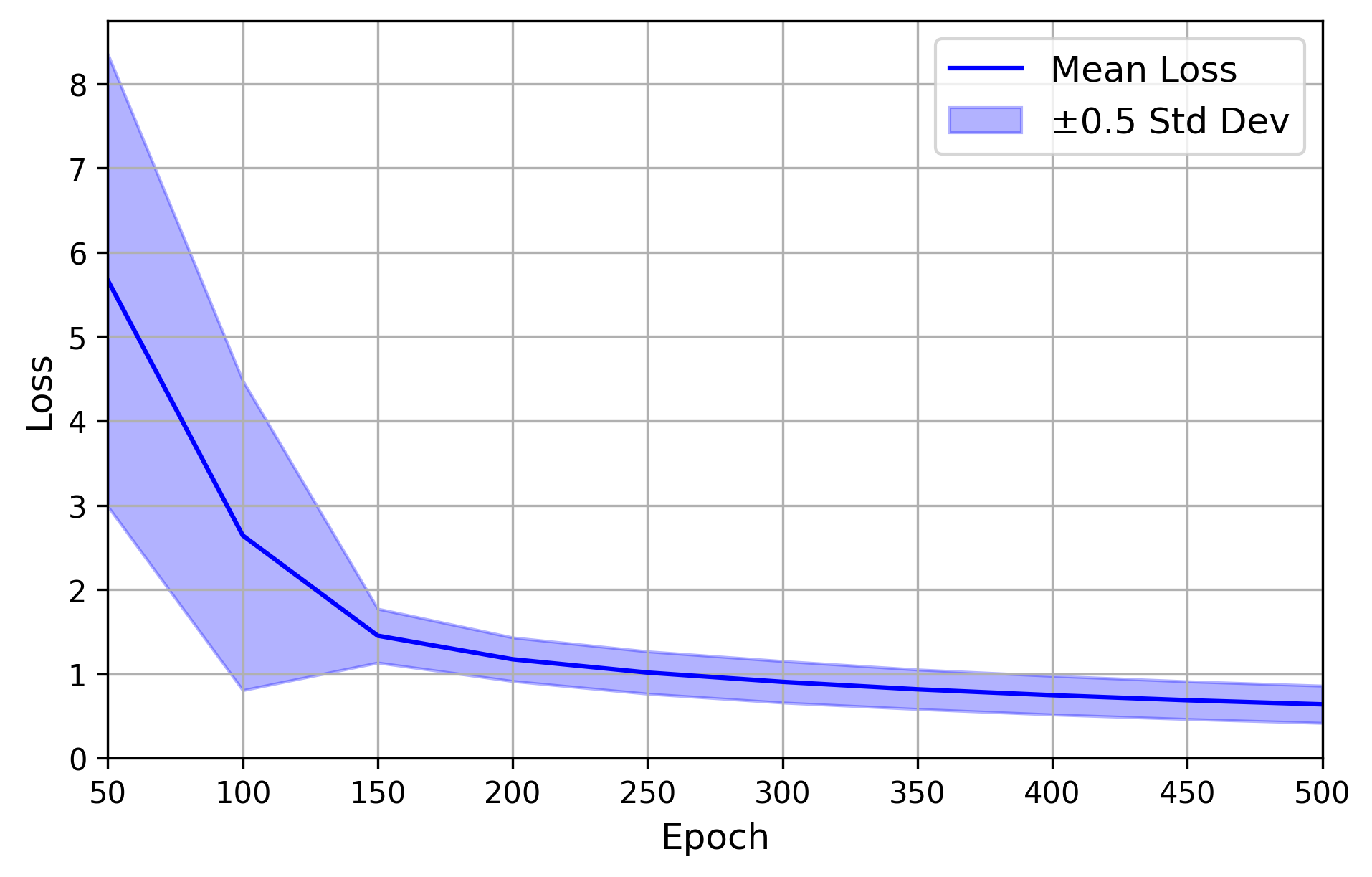}%
}\hfil
\subfloat[client-side DP\label{fig:withLDPLV}]{%
  \includegraphics[width=8cm, height=5.5cm]{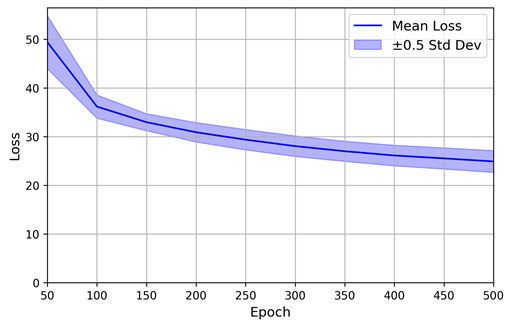}%
}
\caption{The smoothed line represents the mean latent vector loss over epochs, with the shaded region indicating $\pm 0.5$ standard deviation. Under client-side DP, the loss remains consistently higher than the non-DP setting, thereby limiting the DCGAN's ability to effectively reconstruct class specific images.
}
    \label{fig:LVectorLoss}
\end{figure*}

\begin{figure*}[!t]
    \centering
    \includegraphics[width=15cm, height=7.5cm]{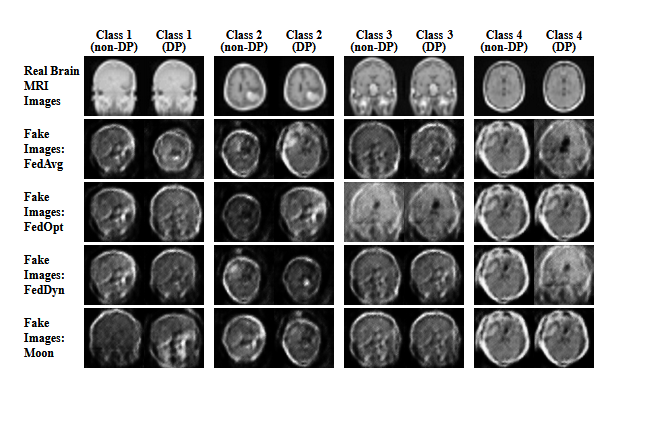}
    \caption{Comparison of real and DCGAN-generated images. In the non-DP setting, clean feedback enables the generation of accurate class-specific samples. In contrast, client-side DP noise distorts the feedback signal, leading to noticeable degradation in fine-grained tumor features.}
    \label{fig:image-recovery}
\end{figure*}

\subsubsection{DCGAN architecture} \label{DCGANArchitecture}
DCGAN comprises two networks: the \textit{generator} and the \textit{discriminator}. Table \ref{tab:generator} and Table~\ref{tab:discriminator} illustrate the architectures of both networks used for DCGAN-based attacks. The \textit{generator} transforms a latent vector of shape  $(1024, 4, 4)$ into a $64 \times 64$ grayscale image using four \texttt{ConvTranspose2D} layers with stride 2 and padding 1. Each layer increases spatial resolution and reduces channel depth, followed by \texttt{InstanceNorm2d} and \texttt{ReLU}, with a final \texttt{Tanh} activation to produce normalized pixel values in [-1, 1]. The \textit{discriminator} processes $64 \times 64$ images through a sequence of \texttt{Conv2D} layers with stride 2, halving the spatial size at each step. The first layer uses \texttt{LeakyReLU}, while subsequent layers use \texttt{BatchNorm2d} and \texttt{LeakyReLU}. The final layer applies a \texttt{Sigmoid} function to produce a scalar output indicating the probability that the input image is real. This adversarial setup enables discriminator to distinguish real images from those generated by the generator.

\subsubsection{Train DCGAN by auxiliary data} \label{DCGANtrain}
Using the auxiliary dataset, we trained the DCGAN for 300 epochs. During training, we observed that the discriminator loss steadily decreased, indicating improved capability in distinguishing between real and fake images. In contrast, the generator loss oscillated and showed a slight upward trend over time. This behavior is expected: as the generator improves and produces higher-quality images, the discriminator also becomes more accurate, making it harder for the generator to deceive it. Consequently, the generator loss may increase slightly, even as image quality improves--reflecting the adversarial balance between the two networks. 
Figure~\ref{fig:MRI} illustrates representative original brain MRI images from each of the four classes, while Figure~\ref{fig:SynMRI} shows synthetic brain MRI images generated by trained DCGAN. While the generated images may appear realistic, they are insufficient for reliably inferring specific class membership. This observation is further supported by the results presented in the privacy evaluation section. Moreover, since no model outputs are queried by the attacker, the generated synthetic images by DCGAN generator were not class specific.

\subsubsection{Latent vector optimization}
For each class, our goal is to generate a set of optimized latent vectors that can produce synthetic MRI images resemble the corresponding target class. To accomplish this, we initialize random latent vectors of shape \((1024, 4, 4)\) for both non-DP and client-side DP settings and iteratively refine them through an optimization process over multiple epochs.

Figure~\ref{fig:withoutLDPLV} shows the mean loss curve, along with a shaded band representing the $\pm0.5$ standard deviation, across 20 latent vector optimization runs under the non-DP settings. Each run yields an optimized latent vector for a specific target class. As training progresses (x-axis: epochs), the average loss steadily decreases, indicating that the generated images increasingly resemble the target class characteristics. The shaded region reflects variability in loss across different runs, which arises due to random initialization and the non-convex nature of the optimization landscape. Some latent vectors begin closer to optimal representations and converge faster, while others require more iterations or settle into suboptimal solutions.

Figure~\ref{fig:withLDPLV} illustrates the same optimization process in the client-side DP setting. While the mean loss also decreases steadily, the final loss values remain significantly higher than in the non-DP case, and the variance is substantially larger. This suggests that differential privacy introduces noise that impairs the optimization process, making it more difficult for the latent vectors to converge to accurate class-specific representations. While convergence continues, it indicates that reconstruction is now harder to accomplish with precision. These results demonstrate that while DP significantly reduces the effectiveness of model inversion attacks, it does not entirely eliminate them. This highlights a fundamental trade-off between privacy and accuracy in FL systems under adversarial conditions.

\begin{figure}[!t]
   \centering
    \includegraphics[width=8cm, height=5.5cm]{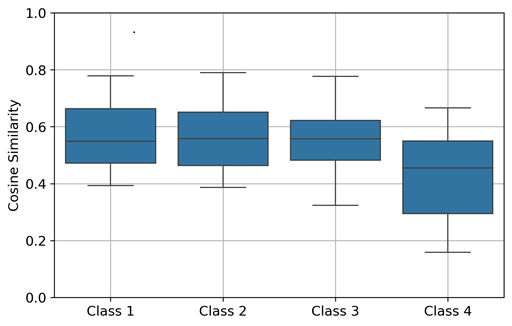}
    \caption{Boxplot of cosine similarity scores between DCGAN-generated and real MRI images without client-deployed global model feedback, showing limited class-specific alignment.}
    \label{fig:cosinesimilarityclassless}
\end{figure}

\subsubsection{Recover class-specific image}
Figure~\ref{fig:image-recovery} provides a visual comparison of reconstructed images for a specific target class using four different aggregation algorithms under both non-DP and DP settings. The first row shows real brain MRI images from each class for reference, while remaining rows display synthetic images reconstructed using optimized latent vectors across the different aggregation methods.

In non-DP setting, all algorithms reveal more structural detail and sharper tumor patterns, indicating that the global model provides richer gradient feedback, which facilitates more accurate reconstructions. When client-side DP is applied to the global model, although DCGANs can still generate visually plausible images, reconstructions consistently lack fine-grained features across all methods due to DCGAN generative capability, the reconstructions consistently lack the fine-grained features present in the non-DP setting--reflecting a reduced chance of success for the attacker. This observation is further supported by the results presented in the privacy evaluation section. We observe a similar trend across all four algorithms.

\begin{figure}[!t]
\centering
\subfloat[Without DP\label{fig:withoutLDPloss2}]{%
  \includegraphics[width=8cm, height=5.5cm]{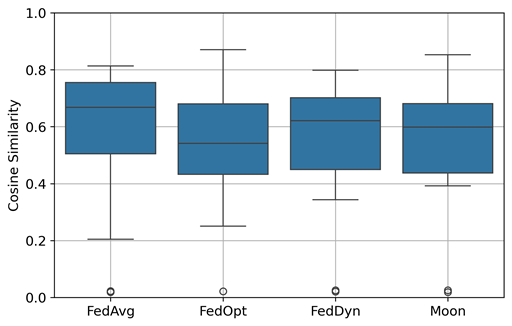}%
}\hfil
\subfloat[With client-side DP\label{fig:withLDPloss2}]{%
  \includegraphics[width=8cm, height=5.5cm]{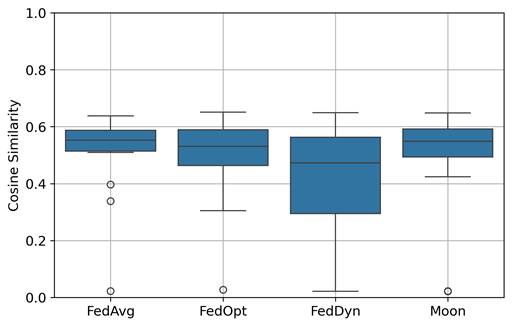}%
}
\caption{Cosine similarity scores using client-deployed global model feedback for (a) non-DP and (b) client-side DP settings. All aggregation algorithms are vulnerable to model inversion attacks in the absence of client-side DP. The use of DP reduces the effectiveness of model inversion attacks.
}
    \label{fig:cosinesimilarity}
\end{figure}

\subsection{Privacy Evaluation}
This section examines the privacy risks from DCGAN-based model inversion attacks in FL. We evaluated our DCGAN-based attack by measuring the cosine similarity between feature vectors of real and reconstructed MRI images. The evaluation is conducted for three types of reconstructed synthetic images: (i) generated by a pre-trained DCGAN without client-deployed trained global model feedback, (ii) generated with global feedback under non-DP setting, and (iii) generated with global feedback under client-side DP setting. In all cases, a pruned version of the pre-trained global model $\phi(\cdot)$ with the final classification layer removed was used as a feature extractor for both real and synthetic MRI images.

\begin{figure*}[!t]
\centering
\subfloat[FedAvg under client-side DP\label{fig:withoutLDPloss1}]{%
  \includegraphics[width=8cm, height=5.5cm]{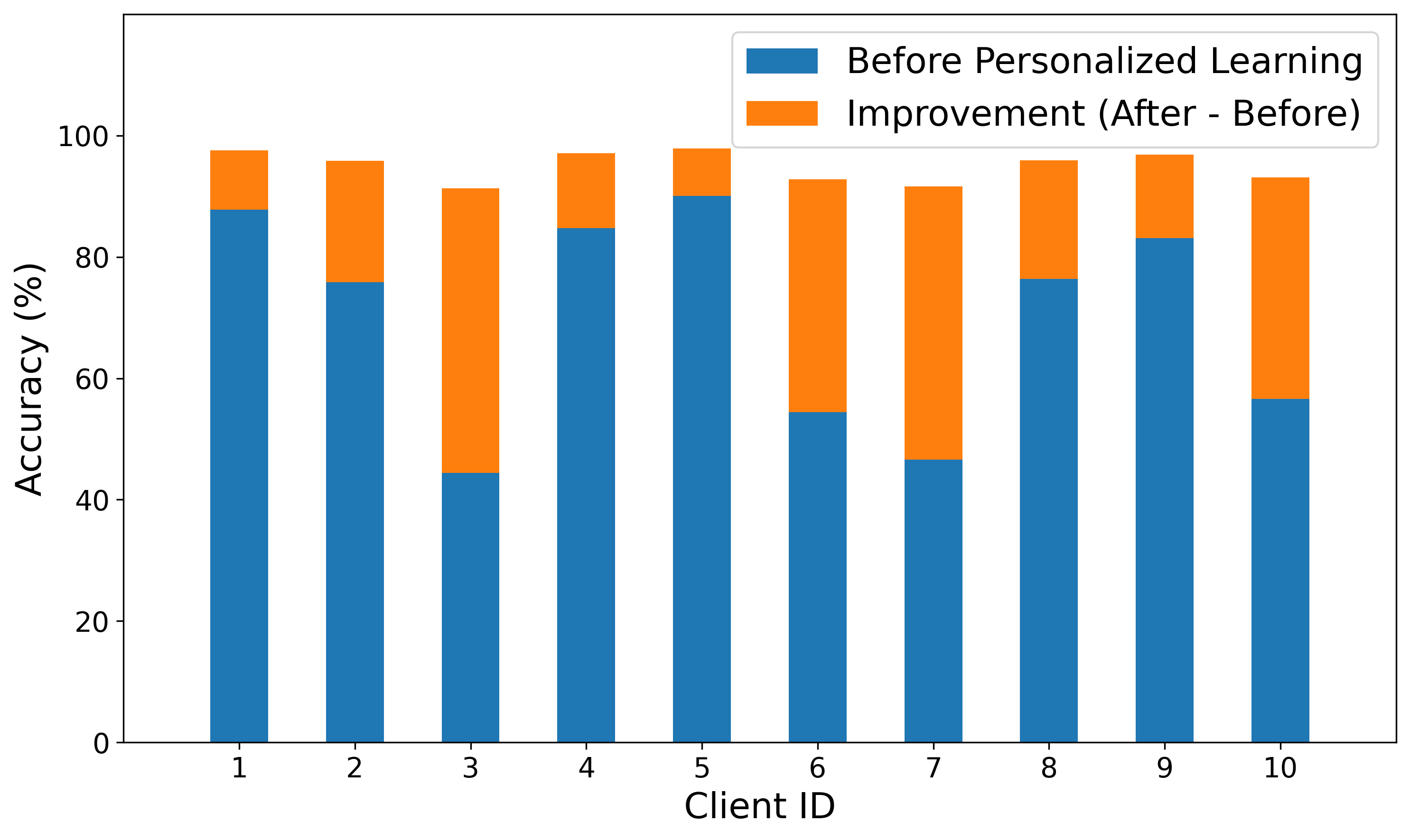}%
}\hfil
\subfloat[FedOpt under client-side DP\label{fig:withLDPloss1}]{%
  \includegraphics[width=8cm, height=5.5cm]{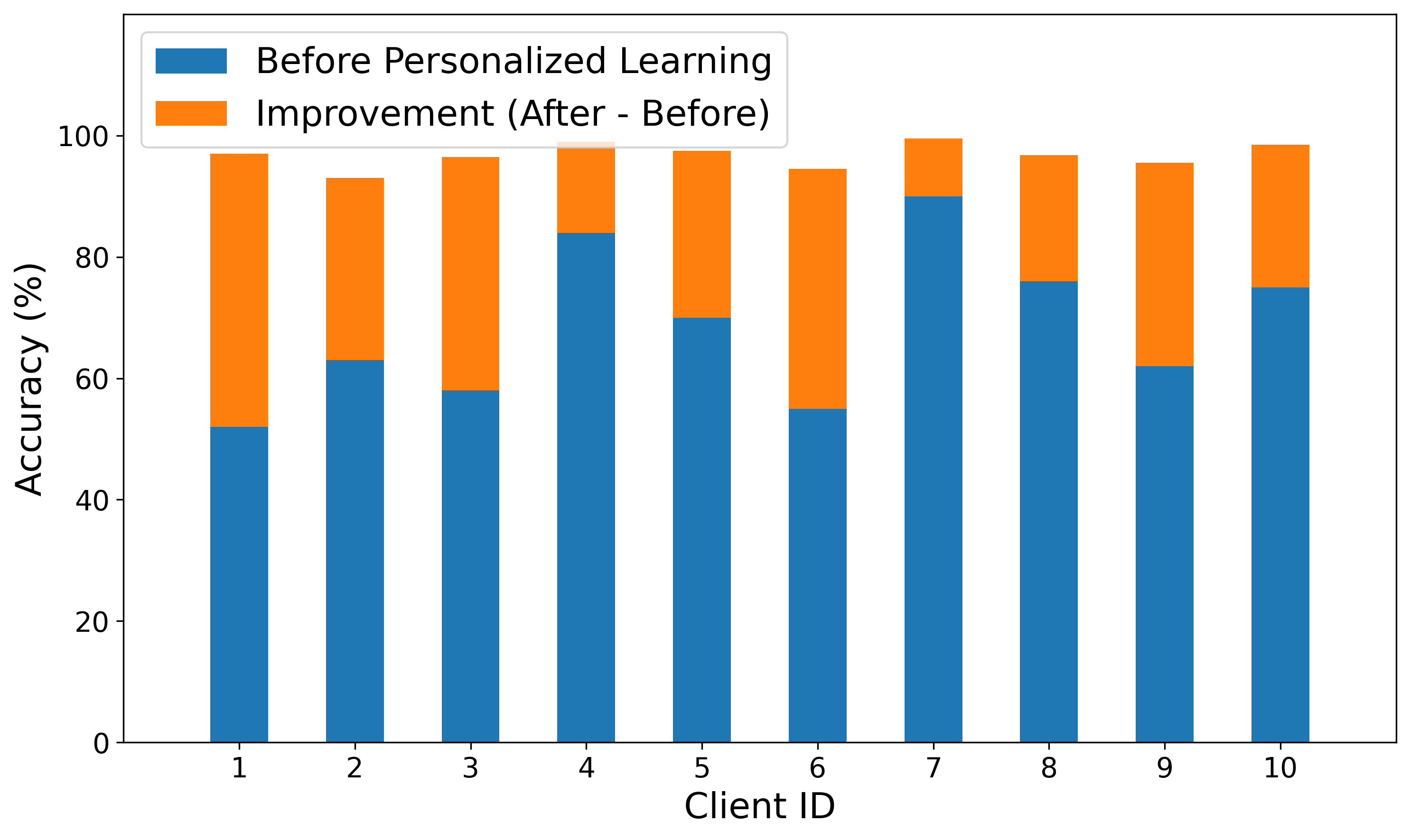}%
}
\caption{Personalization applied to the pre-trained client-deployed global models trained with (a) FedAvg under client-side DP and (b) FedOpt under client-side DP. In both cases, personalization leads to significant improvements in client-level accuracy. 
}
    \label{fig:personalization}
\end{figure*}

\subsubsection{Real vs. synthetic MRI image without global model feedback} \label{similaritywithourfeedback}
In the absence of client-deployed global model feedback, the pre-trained DCGAN operates solely based on its inherent generative capability, which primarily depends on the available auxiliary data. As shown in Figure~\ref{fig:cosinesimilarityclassless}, the resulting cosine similarity scores range from approximately 0.15 to 0.79 across the four tumor classes. Higher similarity scores in some cases suggest that the auxiliary dataset includes images resembling those in the training data. Overall, the distributions indicate moderate reconstruction quality, with noticeable variation across all classes. For instance, Class~2 shows a higher median similarity and a tighter distribution, implying better alignment of auxiliary data with the real data. In contrast, Class~4 exhibits the lowest similarity and a broader spread, indicating weaker reconstruction fidelity. Notably, in this setting, the FL aggregation algorithm has no effect, as no client-deployed global model feedback is utilized.

\subsubsection{Real vs. synthetic MRI image with global model feedback} 
Figure~\ref{fig:cosinesimilarity}(a, b) presents box plots comparing the cosine similarity distributions of the reconstructed images for the first tumor class across four aggregation algorithms under both non-DP and client-side DP settings.

In the non-DP setting, FedOpt exhibits the highest similarity score (0.87), indicating significant privacy leakage in some cases, although its wide distribution reflects inconsistency. FedAvg also shows high similarity scores (up to 0.81) but with greater variability, including some very low values. FedDyn demonstrates more stable, moderate similarity values (mostly between 0.55 to 0.73). MOON shows a similar pattern but includes an outlier reaching 0.85.

Under the client-side DP setting, the effectiveness of the model inversion attack significantly diminishes. Cosine similarity scores are generally lower, indicating that DP noise disrupts the feedback used for reconstruction. FedAvg still yields moderately high values (up to 0.64), though many samples fall below 0.55. FedOpt and MOON show reduced and more consistent scores, mostly ranging between 0.5 to 0.65. FedDyn exhibits the most notable privacy improvement, with many values below 0.45 and some as low as 0.3-demonstrating strong resistance to model inversion under client-side DP.

These results confirm that while none of the aggregation algorithms can fully prevent semantic reconstruction in the absence of client-side DP, the introduction of DP significantly reduces information leakage—an observation consistent with the visual comparisons in Figure~\ref{fig:image-recovery}—although this comes at the cost of reduced classification accuracy. Similar results are observed for other aggregation algorithms as well.

\subsection{Personalized Federated Learning}
In the personalized experiments in PPP-FL, we focused on cases where a client-deployed trained global model performs poorly under the client-side DP setting and examined how personalization can improve client-level accuracy. To do so, we first selected a global model trained with client-side DP for a particular aggregation algorithm that exhibited suboptimal performance. We then applied personalization to this global model using each client's local data.

Figure~\ref{fig:personalization}(a) illustrates the effect of \textit{personalized FL} for FedAvg under the DP setting with 10 clients. The baseline global model accuracy averaged 68.9\% across all clients. The blue bars represent each client’s accuracy before personalization, while the orange bars show the accuracy after personalization, which increases to an average of 95.88\% when using a regularization parameter of $\lambda = 1 \times 10^{-3}$. A larger value of $\lambda$ encourages stronger alignment with the global model, whereas a smaller value places more emphasis on local personalization. Most clients experience substantial accuracy improvements—often approaching the performance of a centralized ML model trained on the full dataset. However, the gains are not entirely uniform due to variations in local dataset sizes and label distributions, and some residual bias persists under highly non-IID conditions.

A similar pattern is observed for \textit{personalized FL} with FedOpt under the client-side DP setting, as shown in Figure~\ref{fig:personalization}(b). The baseline global accuracy of 68.5\% across 10 clients (blue bars) increases to 96.78\% after personalization (orange bars). Overall, these results clearly demonstrate that personalization substantially improves client-level performance when the global FL model is underperforming, particularly in DP-based training scenarios.

\subsection{Discussion and Limitations}
Our experimental findings suggest that the vulnerability of FL systems to DCGAN-based model inversion attacks is influenced by the informativeness of client-deployed global model feedback, and the quality of auxiliary data. Although different aggregation strategies show minor variations, their influence on reconstruction quality is relatively limited. The information contained in the client-deployed global model’s feedback--particularly in the non-DP setting--guides the optimization of latent vectors, enabling the attacker to refine synthetic data toward class-specific representations. Additionally, when the auxiliary dataset is drawn from a domain similar to the training data, the DCGAN can generate structurally realistic samples even before receiving feedback, further improving the feature reconstruction quality. In contrast, client-side DP significantly disrupts the optimization process by injecting noise, thereby reducing information leakage and weakening semantic reconstruction. Overall, these results highlight the fundamental trade-off between privacy and utility in FL systems, as stronger privacy protections often limit reconstruction but also reduce model accuracy. Personalization within FL substantially enhances local model performance under client-side DP, helping clients recover much of the accuracy lost due to privacy-preserving noise.

A key limitation of this study is its reliance on auxiliary data that closely resemble the training distribution, while in practice attackers may not have access to such representative datasets. It is also important to evaluate more realistic scenarios involving large, high-resolution MRI images with variable resolutions and non-IID data distributions, which can affect global model convergence. In addition, loss-based feedback may not always be accessible to the attacker, further limiting the practicality of model inversion attacks. Future work will also extend this analysis to a wider range of FL aggregation strategies.

\section{Conclusion} \label{sec:conclusion}
This study investigates privacy risks in federated learning systems under loss-based black-box model inversion attacks. We propose a practical, model-agnostic attack framework using DCGANs that leverages unlabeled auxiliary data and client-deployed trained global model feedback to reconstruct class-specific MRI images. Our results show that the success of the attacker depends more on the informativeness of the feedback and the similarity of the auxiliary data than on the aggregation strategy. In non-DP settings, all tested algorithms leak semantic features, whereas applying client-side DP significantly weakens the attack by disrupting latent vector optimization and reducing reconstruction quality. Importantly, model personalization substantially enhances local performance—particularly in client-side DP-enabled settings—ensuring that each client benefits from context-specific model refinement. These results confirm that personalized FL helps recover model utility under DP constraints while privacy guarantees are enforced by the DP mechanism. Future work will explore broader data domains, weaker auxiliary priors, and stronger defense mechanisms under heterogeneous conditions.

\section*{References}

\begin{IEEEbiography}%
[{\includegraphics[width=1in,height=1.25in,clip,keepaspectratio]{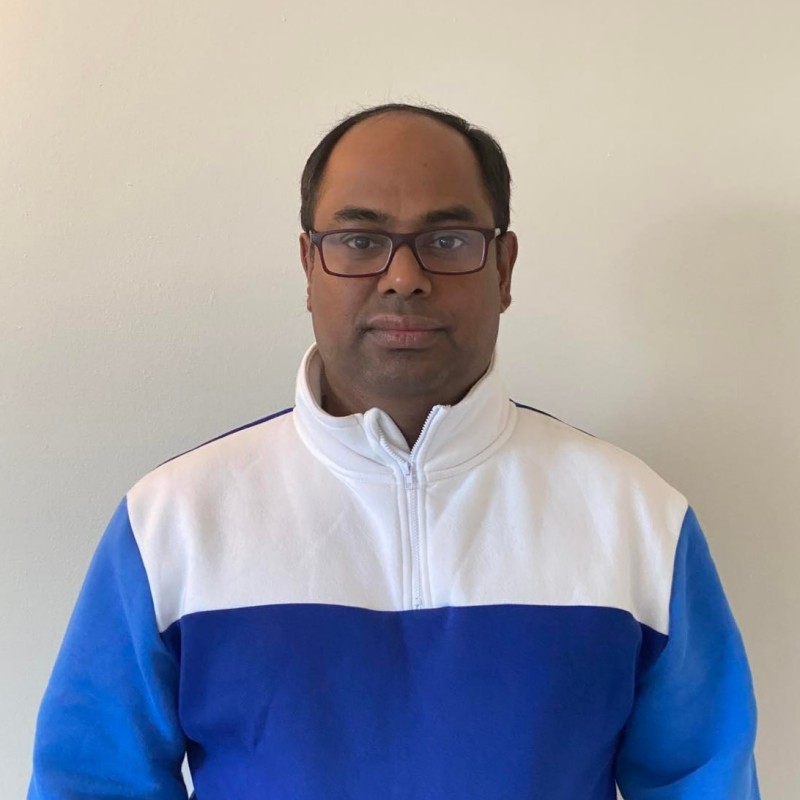}}]
{Dr. Md Morshedul Islam} is currently a full-time Assistant Professor in the Department of Mathematics and Information Technology at Concordia University of Edmonton (CUE). He received his Ph.D. degree in computer science from the University of Calgary (UofC), where his research focused on data privacy and security. He previously worked as a Postdoctoral Fellow in the Department of Electrical and Software Engineering at the UofCs. He has authored and co-authored several peer-reviewed journal articles and conference papers in venues such as IEEE Access, ACM CCS, FPS, IFIP SEC, and SECRYPT. His research interests include data privacy and security, machine learning, and responsible AI. Dr. Islam is actively involved in supervising graduate and undergraduate students, managing research grants, and engaging in international research collaborations.
\end{IEEEbiography}
\vspace{-5mm}
\begin{IEEEbiography}%
[{\includegraphics[width=1in,height=1.25in,clip,keepaspectratio]{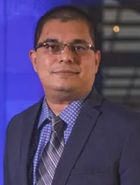}}]
{Dr. Hossain Shahriar} is Associate Director and Professor for the Center for Cybersecurity at the University of West Florida. His research interests include mobile and web security, mhealth, EHR systems and healthcare security, HIPAA compliance checking, malware analysis, automatic checking of vulnerabilities, and mitigation. His research projects have been supported by the National Science Foundation, National Security Agency, the Department of Defense, National Institute of Health, and private industry partners. He developed and taught courses with open-source materials such as Ethical Hacking, AI in Cybersecurity, Health Information Security and Privacy, Computing Infrastructure, and Information Security Concepts and Administration. Dr. Shahriar has served as Program Chair (IEEE ICDH), Publication Chair (ACM SAC), and Proceedings Chair (IEEE COMPSAC)  in conferences.
\end{IEEEbiography}
\vspace{-5mm}
\begin{IEEEbiography}%
[{\includegraphics[width=1in,height=1.25in,clip,keepaspectratio]{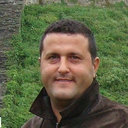}}]
{Alfredo Cuzzocrea} is Distinguished Professor of Computer Engineering and the Founder and Director of The Big Data Engineering and Analytics Laboratory at the University of Calabria, Rende, Italy. He is also Research Associate of the National Research Council (CNR), Rome, Italy. His current research interests span the following scientific fields: big data, database systems, OLAP, data warehousing, and knowledge discovery. He is author or co-author of more than 830 papers in international conferences, journals and books.
\end{IEEEbiography}
\end{document}